\documentstyle[12pt,cite,epsf]{article}                                        
\catcode`\@=11                                                       
\begin{document}                                                     

\renewcommand{\theequation}{\thesection.\arabic{equation}}        
\newcommand{\mysection}[1]{\setcounter{equation}{0}\section{#1}}

\def\1{{\bf 1}}
\def\Z{{\bf Z}}
\def\ee{\end{equation}}
\def\be{\begin{equation}}
\def\la{\label}
\def\dxi{\partial_\xi}
\def\D{{\cal D}}
\def\sin{{\rm sin}}
\def\cos{{\rm cos}}
\def\f{{\bf \Phi}}
\def\v{\varphi}
\def\O{\bf {\cal O}}
\def\C{\bf C}
\def\T{\Omega}
\def\CP{\bf CP}
\def\e{\rm e}
\def\0{\nonumber}
\def\eea{\end{eqnarray}}
\def\bea{\begin{eqnarray}}
\def\Tr{\rm Tr}
\def\IR{\bf R}
\def\ZZ{\bf Z}

%
%
\input epsf
\newcounter{fignum}
\newcommand{\figuurnum}{\arabic{fignum}}
\newcommand{\figuur}[2]{
\addtocounter{fignum}{1}
\addcontentsline{lof}{figure}{\protect
\numberline{\arabic{section}.\arabic{fignum}}{#2}}
\hspace{-3mm}{\it fig.}\ \figuurnum.
\begin{figure}[t]\begin{center}
\leavevmode\hbox{\epsffile{#1.eps}}\\[3mm]
\parbox{10cm}{\small \bf Fig.\ \figuurnum : \it #2}
\end{center} \end{figure}\hspace{-1.5mm}}
%
%
\newcommand{\figuurplus}[3]{
\addtocounter{fignum}{1}
\addcontentsline{lof}{figure}{\protect
\numberline{\arabic{section}.\arabic{fignum}}{#3}}
\hspace{-3mm}{\it Fig.}\ \figuurnum.
\begin{figure}[t]\begin{center}
\leavevmode\hbox{\epsfxsize=#2 \epsffile{#1.eps}}\\[5mm]
\parbox{10cm}{\small \bf Fig.\ \figuurnum: \it #3}
\end{center} \end{figure}\hspace{-1.5mm}}
\newcommand{\fig}{{\it fig.}\ }


\begin{titlepage}

\hfill{ULB-TH/01-35}

\vspace{2 cm}

\centerline{{\huge{The M5-brane on K3 and del Pezzo's}}}
\centerline{{\huge{and multi-loop string amplitudes}}} 

\vspace{2 cm}

\centerline{Giulio Bonelli
\footnote{e-mail bonelli@phys.uu.nl}}
\centerline{Physique Theorique et Mathematique - Universite Libre de Bruxelles}
\centerline{Campus Plaine C.P. 231 - B 1050 Bruxelles, Belgium}
\centerline{and}
\centerline{Theoretische Natuurkunde - Vrije Universiteit Brussel}
\centerline{Pleinlaan, 2 - B 1050 Brussel, Belgium}

\vspace{4 cm}

{\it Abstract}:
We study the BPS spectrum of Little String Theory for bound states 
of M5-branes wrapped on six manifold of product topology
$M_4\times\Sigma_2$ and the apparence of multi-loop $\theta$-functions
in a supersymmetric index calculation. 
We find a total reconstruction of the g-loop heterotic contribution 
in the case of a double K3 M-theory compactification.
Moreover, we consider total wrapping of M5-branes on del Pezzo
surfaces $B_k$ and, by studying the relevant amplitude, we notice 
the arising of $\theta$-functions relative to BPS strings on $T^{k-1}$,
i.e. membranes on $T^k$.
This happens because of beautiful relations between four dimensional SYM theories 
and CFTs in two dimensions and seems to be linked to a duality
recently observed by A.Iqbal, A.Neitzke and C.Vafa in \cite{inv}.

\end{titlepage}

\section{Introduction}

The problem of the world-volume theory of the 5-brane in M-theory is still an open one
and its solution is one of the major points still awaiting for a clarification 
in order to have a clearer picture of non-perturbative string theory.

There exists \cite{open} a conjectural solution of this problem in terms of a six-dimensional
string theory, called Little String Theory, which should describe the
world-volume theory of the M5-brane. This arises naturally once we understand 
the M-theory 5-brane both as the magnetic dual of the membrane and as Dirichlet
surface for membrane's boundaries. Notice that this picture is natural
from the point of view of the equivalence of M-theory compactification on a circle
$S^1$ and type IIA string theory. In fact, under compactification the M5-brane
has to generate both NS-5branes which are magnetic duals to fundamental strings
and D4-branes on which fundamental strings can end. At the same time the fundamental 
string is generated by membranes wrapped on the circle.

In this paper we continue the analysis of an on-shell model for LST
which was initiated in \cite{io} where a compactification of M-theory
on $CY_6\times T^2$ was considered and a six dimensional scheme 
for computing the supersymmetric index of N 5-branes wrapped on
$M\times T^2$, where $M$ is a four dimensional supersymmetric cycle in 
$CY_6$, was given. Moreover, in \cite{io}, a natural generalization 
for the supersymmetric index was given.

The scheme which has been proposed in \cite{io} is the following.
As far as the {\it single 5-brane} is concerned, 
we obtained that the supersymmetric index
of the world-volume theory is made from two multiplicative factors.

One corresponds to the self-dual two tensor which describes the 5-brane 
zero-modes with respect to 11-dimensional supergravity or, equivalently
\cite{bb}, the 11-dimensional membrane theory
and is calculated as a zero characteristic
generalized $\theta$-function as given 
in \cite{mans}. This depends upon the six manifold where the 5-brane is wrapped as
$$
\theta\left[\matrix{\alpha \cr \beta}\right](hZ^0_{\cal C}|0)
=\sum_k e^{i\pi \left((k+\alpha)Z^0(k+\alpha)+2(k+\alpha)\beta\right)},
$$ where $Z^0$ is the period
matrix of the relevant six-manifold cohomology that is specified shortly.
Let $\left\{E^{(6)},\tilde E^{(6)}\right\}$ be a symplectic basis
of harmonic 3-forms on the six-manifold at hand such that, in matrix notation,
$$
\int E^{(6)}E^{(6)}=0,\quad
\int \tilde E^{(6)}E^{(6)}=1,\quad
\int \tilde E^{(6)}\tilde E^{(6)}=0.
$$
We can expand $\tilde E^{(6)}=X^0E^{(6)}+Y^0{}^*E^{(6)}$, where ${}^*$ is the Hodge
$*$-operator. Then $Z^0$ is defined as $Z^0=X^0+iY^0$.

A second multiplicative factor comes from Little String states and  
depends again upon the moduli of the world-volume manifold.
In principle it is given by the supersymmetric index of the Little String
Theory on the six dimensional world-volume and hence by a one-loop
path-integral in a twisted supersymmetric $\sigma$-model.

Therefore the total index formula is
\be
{\cal I}_1=[LST]_{{\rm one\,\, loop}}\times\Theta(Z^0)
\label{general1}\ee

Here it comes the point of interest. In fact we don't know what is the actual
formulation of LST and therefore we don't have still a general rule for the calculation 
of this factor and indeed the point is to find possible tools to infer
its value. A possibility, which was successfully explored in \cite{io}, is to 
use super-string duality to map the counting of BPS multiplets
to a better understood sector of the theory and exploit the calculation there.
Then, inverting the duality map, we can give a reinterpretation of the results 
to get some news about the actual structure of LST.
Actually the off-shell model proposed in \cite{io} assumes that there exists
a single type of saturated LST multiplets which has to be token into account.
This is enforced by the analysis of $(0,2)$ short multiplets carried out in \cite{mans2}
where a single type of these objects was found.

In the case of {\it multi 5-branes} bound states the calculations in \cite{io}
suggest a general formula for the supersymmetric index which goes as follows.
Suppose we are dealing with a BPS bound state of $N$ 5-branes wrapped on
a six manifold {\cal C}. Then we have to consider the space of 
rank N holomorphic coverings of {\cal C} in the ambient space
of M-theory. Generically this space will be disconnected and reducible 
into components consisting of connected irreducible coverings
of rank less or equal to $N$. Each of these irreducible holomorphic 
coverings
corresponds to an irreducible bound state.
Then the general structure formula for the supersymmetric index
of an irreducible bound state of $N$ 5-branes is given by the lifting to the spectral 
cover of the index formula.
In formulas, if $C_N$ is a generic holomorphic covering of $C$, then $C_N=\cup_j C_j^{irr}$,
where $\sum j=N$ and $C_j^{irr}$ is irreducible, and the relative contribution to the index 
is ${\cal I}_N^C=\prod_j {\cal I}_1^{C_j^{irr}}$.

The aim of this paper is to focus on the above $\Theta$-function contribution.
We consider 5-branes wrapped on six manifolds with product topologies $M_4\times\Sigma_2$.
In section 2 we calculate the $\Theta$-function contribution in general and we show 
that it fits within a path-integral calculation for $U(1)^g$
gauge theories by identifying the coupling matrix with the Riemann surface period matrix.
Then we move to particular cases. In section 3 we review the case of total wrapping on $K3\times T^2$
by showing its natural interpretation as a one loop heterotic string amplitude on $T^3$ and 
in section 4 we consider more complicated geometries $K3\times \Sigma$ 
which give the multiloop heterotic $\theta$-function amplitudes as a result.
In section 5 we study the case in which $M_4$ is a del Pezzo surface $B_k$ and 
expose the multiloop $\theta$-function corresponding to 1/2 BPS strings on $T^{k-1}$,
i.e. membranes in M-theory on $T^k$ and try some connections with
the geometric duality proposed in \cite{inv}.
Some other issues and open questions are discussed in the last section.
The Appendix contains the calculation of the period matrix for six manifolds 
of the above product topology.

\section{$\theta$-functions and 5-branes}

The aim of this section is to calculate the partition function of $g$ 
copies of the maximally supersymmetric Maxwell theory on a four manifold $M$  
and to show
that it reproduces the typical string g-loop $\theta$-function
on a genus $g$ Riemann surface $\Sigma$
where the lattice sum is along the lattice of integer period elements in $H^2(M,\IR)$
and the $g\times g$ period matrix is given by the matrix of the couplings
for the theory with $U(1)^g$ gauge symmetry.
We generalize the calculational techniques developed in \cite{witten95}.
Then we compare the result with the calculation of the relevant 
zeromodes contribution coming from the $(0,2)$ tensor multiplet 
on $\Sigma\times M$, where $\Sigma$ and $M$ are as above.

We deal with a set of $g$ Abelian vector multiplets $V^I$, where $I=1,\dots,g$
whose action is
\be
S= I(gauge\, vectors) + I(other\, bosons) +I(fermions)
\label{action}\ee
where
the action $I(gauge)$ is taken to be
$$
I=\frac{i}{4\pi}\sum_{I,J}\int_M \left[\bar \T_{IJ}(F^I)^+\cdot(F^J)^+
- \T_{IJ}(F^I)^-\cdot(F^J)^-\right]
$$
where $\T$ is a complex symmetric $g\times g$ matrix with positive 
imaginary part,
$F^I=dA^I$ are the curvature fields and $(F^I)^\pm$ denotes self dual and anti-self dual
parts of the curvature fields.

The rest of the action in (\ref{action}) is fixed by supersymmetry.
If we consider the ordinary supersymmetric theory (six scalar bosons),
supersymmetry requires the manifold $M$ to be Ricci flat. We will consider
actually a twisted version of the supersymmetric theory which is the Abelian 
analog of that considered in \cite{vafawitten} (three scalar bosons and a self-dual 2-form)
and we will assume the manifold $M$ to be Kahler and 
fulfilling appropriate vanishing theorems.
In both the cases, in calculating the partition function, the integration over the fluctuations 
of the fields gives a constant factor. By itself this factor would be zero, 
because of the presence of 
fermionic zero modes which cause the path-integral to vanish.
To obtain a non zero result, we have to insert fermion fields in the path-integral
in the minimal sufficient amount to soak-up the zero modes integration.
Under this condition the oscillating part of the path integral is a non zero constant
in both the cases. Notice that the Ricci flatness condition 
(which would be relevant for the ordinary supersymmetric case)
implies a reduction of the 
holonomy group of the manifold $SO(4)\to SU(2)$ and this is the case when the 
spin relabeling of the fields, which relates the
twisted and the ordinary supersymmetric theory,
is just a change of variables since the differently coupled part of the spin connection
(which in principle would make the theories inequivalent) is null.

The zero modes of the curvature fields span the lattice $\Lambda_{b_+,b_-}$ of 
elements of $H^2(M,\IR)$ with integer periods as $F^I=m^I/2\pi$, $m^I\in\Lambda_{b_+,b_-}$.
Each one of them can be uniquely decomposed in a self dual and an anti-self dual part
as $m=\frac{1}{2}(m+{}^*m)+\frac{1}{2}(m-{}^*m)$.

Therefore we evaluate easily the zero mode part of the partition function
in terms of the intersection matrix $Q$ on $\Lambda_{b_+,b_-}$ as
$$
{\cal Z}_{zm}^M=
\sum_{m^I\in\Lambda_{b_+,b_-}}e^{-I(m)}=
\sum_{(k^I_R,k^I_L)\in\Gamma_{b_+,b_-}}
e^{i\pi k^I_L \T_{IJ} k^J_L -i\pi k^I_R \bar \T_{IJ} k^J_R}
\equiv \Theta_{\Gamma_{b_+,b_-}}(\Omega)
$$
where $k^I_R=\frac{\1 + Q}{2}k^I$
and $k^I_L=\frac{\1 - Q}{2}k^I$ and $k^I\in\ZZ^{b_2}$ for $I=1,\dots,g$.

This is the multiloop $\theta$ function for the lattice $\Gamma_{b_+,b_-}$
and its appearing here is a sign of a very deep link between string theory
and four dimensional gauge theories. This connection is in fact due to the 
following remarkable coincidence.
Indeed, we 
notice that exactly the same multi-loop $\theta$-function arises from the 
calculation of the zero-modes contribution of the self-dual tensor
multiplet on a six manifold $\Sigma\times M$, where $\Sigma$ is a genus 
$g$ Riemann surface with period matrix $\Omega$.

In Appendix A we calculate the period matrix of a generic product topology 
as above and it is given by (see eq.(\ref{perma}))
$$
Z^0=i\1_{b_1}\oplus \left\{
-Q\otimes\Omega^{(1)}+i\1_{b_2}\otimes \Omega^{(2)}
\right\}
$$
where $b_i={\rm dim} H^i(M,\IR)$, $Q$ is the intersection matrix on $H^2(M,\Z)$
and $\Omega=\Omega^{(1)}+i\Omega^{(2)}$.
By direct calculation we find
\be
\theta\left[\matrix{\alpha=0 \cr \beta=0}\right](Z^0|0)
= \sum_{k\in\ZZ^{b_1+gb_2}}e^{i\pi k\Z_0k}
= k^{b_1}
\times \Theta_{\Gamma_{b_+,b_-}}(\Omega)
\ee
where $k=\theta_3(0|i)=\sum_{n\in\ZZ} e^{-\pi n^2}=1,086434...$ is an
irrational number.

Going back to the calculation of the supersymmetric index of the four dimensional gauge theory, 
there is a further multiplicative term 
which is due to point-like instantons.
Generalizing the techniques developed in \cite{vafawitten} to our
case we get
\be
\prod_{I,J}\sum_n (q_{IJ})^{(n-\chi/24)}\chi(M^n/S_n)=\prod_{IJ}\left[\eta(q_{IJ})\right]^{-\chi}
\label{point}\ee
where $\eta$ is the Dedekind $\eta$ function and $q_{IJ}=e^{2\pi i \T_{IJ}}$.
One can check the modular properties of the above functions via Dehn 
twists techniques, following for example the discussion in \cite{GV}.

\section{Total wrapping on K3}

A first case of interest is when the M5-brane wraps an entire $K3$ and a $T^2$
in M-theory. We include it here for completeness, although
these results already appeared in \cite{Estrings,io2}, to introduce the M-theory/Heterotic 
duality that will be used later in the text.

The period matrix can be obtained from the general result 
given in Appendix A and it is given by
$$
Z_0= -Q' \tau^{(1)}+i\1_{22}\tau^{(2)}
$$
where $\tau=\tau^{(1)}+i\tau^{(2)}$ is the modulus of $T^2$
and $Q'$ is the intersection matrix on $K3$, that is 
$$
Q'=\left(\matrix{0 & 1 \cr 1 & 0\cr}\right)^{\oplus 3}\oplus C_{E_8}\oplus C_{E_8}
$$
where $C_{E_8}$ is the $E_8$ Cartan matrix.
In this case we have 
$$
\Theta(Z_0)=|\theta_3(\tau)|^6 (\Theta_{E_8}(\tau))^2
$$
where we used standard $\theta$-function notation ($\theta_3(\tau)=\theta_3(0|\tau)$).

In this case the Little String contribution to the index is also already
calculated in \cite{io} to be 
$\eta(\tau)^{-\chi}$ where $\chi$ is the Euler characteristic of $K3$, that is $\chi=24$.
Therefore, all in all, we find that the supersymmetric index for a 5-brane on $K3\times T^2$
is given by
$$
{\cal I}_1^{K3\times T^2}=\eta(\tau)^{-24}|\theta_3(\tau)|^6 (\theta_{E_8}(\tau))^2
$$

To understand this result we can make use of 
duality with the heterotic string on $T^3$ under which our 5-brane state gets mapped to 
the fundamental heterotic string \cite{m-het}.
The susy index for this string is computable using the methods explored in \cite{narain}
and it is given exactly as
$$
\frac{1}{\eta(\tau)^{24}}\times |\theta_3(\tau)|^6 (\theta_{E_8}(\tau))^2
$$
where $\frac{1}{\eta(\tau)^{24}}$ is the left moving oscillator partition function,
$ |\theta_3(\tau)|^6 $ the factor taking into account the $T^3$ compact space
and $(\theta_{E_8}(\tau))^2$ the $E_8\times E_8$ gauge group structure.

The same result can be obtained also by duality with type IIB on $K3\times T^2\times\IR^3$
where the 5-branes get mapped to D3-branes on $K3$. Using the results of the previous section 
(which in this case coincide with the analysis 
done in \cite{io}), we find very easily again the above results
when now $\frac{1}{\eta(\tau)^{24}}$ comes from pointlike instantons on $K3$
and $ |\theta_3(\tau)|^6 (\theta_{E_8}(\tau))^2 $ is the zero-mode
partition function for the {\cal N}=4 U(1) theory on $K3$.

As far as the multi M5-brane case is concerned we refer the reader to \cite{io2}
for the explicit derivation of the following index formula
$$
{\cal I}_N^{K3\times T^2}= H_N{\cal I}_1^{K3\times T^2}
$$
where $H_N$ is the modular Hecke operator.
This is naturally interpreted as a multi-instanton toric amplitude in the heterotic string.
Notice also that the same results follow from the application of orbifold techniques 
\cite{toru}.

\section{Partial wrapping on K3}

We consider M-theory on $K3\times T^4\times \IR^3$ and we will explore the above 
six dimensional framework for the counting of BPS multiplets of states for 
a single 5-brane 
wrapped on $\Sigma\times T^4$, where $\Sigma$ is a supersymmetric cycle in $K3$.
This means that $\Sigma$ is an holomorphic curve in $K3$.
As far as the calculation of the $\Theta(Z_0)$ contribution is concerned, we 
calculate the period matrix following the general treatment given in Appendix A.
It is given by
$$
Z_0= i\1_4\oplus \left[-Q\otimes\Omega^{(1)}+i\1_6\otimes \Omega^{(2)}\right]
$$
where $\Omega=\Omega^{(1)}+i\Omega^{(2)}$ is the period matrix of $\Sigma$ and 
$Q=\left(\matrix{0 & 1 \cr 1 & 0\cr}\right)^{\oplus 3}$
the intersection form of $T^4$.

It is straightforward now to obtain 
\be
\Theta(Z_0)=k^4\Theta_{\Gamma_{3,3}}(\Omega)\,
\label{star}\ee
and we find that the lattice sum now covers only uncharged states in the
$\Gamma_{3,3}$ sublattice of the heterotic $\Gamma_{3,19}$. This finds 
a natural explanation in the following paragraphs.

Fortunately we can make use of the M-theory on K3/Heterotic on $T^3$ duality
also in this case, since 
our M5-branes on the heterotic side get mapped to heterotic NS5-branes wrapped on a two cycle 
of $T^3$. These appear as 3-branes and are dual to particles in 7 dimensions and
these particles are nothing but strings on $T^3$.
We see in fact all their momentum/winding lattice at arbitrary loop in the above formula.

In the case when the 5-brane wrap a susy cycle in $K3$, we can make use of the second 
above mentioned duality, that is the one with Type IIB superstring on $K3\times T^2$.
Now the M5-branes get mapped to D3-branes wrapping $\Sigma\times T^2$, but we are not 
yet at the end of the story since to obtain a stable result 
-- actually we shrink the $T^4$ in M-theory to zero volume --  
we have to perform two T-dualities along the leftover shrinking $T^2$. 
Therefore we land finally on type IIB on $K3\times \IR^6$ and our five branes 
get mapped to D1-branes on the original holomorphic two-cycle $\Sigma$ in $K3$.
From this dual point of view it is natural to find only the uncharged states 
in the lattice sum (\ref{star}).

We can combine the two cases by considering M-theory on $K3\times K3'$
and the 5-branes on $\Sigma\times K3'$.
For the single 5-brane case we directly generate the multi-loop
heterotic on ${T^3}'$
$\theta$-function summation from the $K3'$ lattice.
In fact the relevant period matrix results to be
$$
Z^0=-Q'\otimes\Omega^{(1)}+i\1_{22}\otimes\Omega^{(2)}
$$
The relative $\theta$-function
can be obtained by specializing the general formula 
in section 2 which gives
$$
\Theta(Z^0)=
\Theta_{\Gamma_{3,19}}(\Omega)
$$
and corresponds to the heterotic multiloop amplitude as calculated in \cite{GV}.

For the multi-brane case the N covering structure 
\cite{HMST} comes from the spectral surface of the relevant $O(2N)$
Hitchin system on $\Sigma$ which describes these BPS strings as D-strings in the $K3$,
while the $K3'$ does not admit non trivial holomorphic branching coverings.
By duality along the $K3$, the 5-branes appear as 3-branes from the 
seven dimensional point of view and their relevant amplitude is given by a 
four dimensional gauge theory which we might compare with similar to
the one studied in section 2.

\section{5-branes on del Pezzo Surfaces}

It appeared recently a curious duality between toroidal compactifications 
of M theory and del Pezzo surfaces \cite{inv}.
The duality has by now no explicit microscopic origin and 
relates seemingly very different objects.
We will show a deep connection between the two which 
gives some hints in the direction of explaining it as related to Little String Theory
on del Pezzo surfaces.

Let us consider a five-brane wrapped on $\Sigma\times B_k$
in a local Calabi-Yau geometry in M-theory,
where $\Sigma$ is a genus $g$ Riemann surface and $B_k$
a del Pezzo obtained by blowing up $k<9$ points on a ${\bf CP}^2$.
By using the results in appendix A and in section 2, we can promptly 
calculate the period matrix of this product six-manifold
from the period matrix $\Omega$ of $\Sigma$ and 
the intersection form on $B_k$ that is the $(k+1)\times(k+1)$
matrix $Q_k={\rm diag}(1,-1,\dots,-1)$.

Then the relevant $\theta$-function is given by
$$
\Theta(\Omega)_{\Gamma_{1,k}}=
\Theta(\Omega)_{\Gamma_{1,1}} \times \Theta(\Omega)_{\Gamma_{0,k-1}}
$$
where we decomposed the lattice $\Gamma_{1,k}$ in 
$\Gamma_{1,1}\oplus\Gamma_{0,k-1}$ along the lattice vector 
corresponding to the canonical class of the del Pezzo $B_k$.

By contracting the latter, we are left with the holomorphic
(i.e. 1/2 BPS by considering the right sector on the groundstate)
$\theta$ function corresponding to multiloop string amplitudes
on $T^{k-1}$. These lifts naturally to M-theory on $T^k$.
Since we interpret our formula as one-loop amplitudes in LST, we see
all the above states contributing to the trace summation and therefore
we can infer their presence in the Hilbert space of the theory.

We see that this makes a natural correspondence between M5-branes on 
the del Pezzo surface $B_k$ and membranes on $T^k$. We take this result as
a hint for a temptative proof of the duality noticed in \cite{inv}.
Notice in fact that, upon F-theory/M-theory duality, the orthogonality condition to the 
canonical class in M-theory is understood as a decompactification limit in 
F-theory/type IIB.

All this suggests that it should exist 
a generalization of the tensionless string theory obtained in 
\cite{wittenK3} as D3-branes wrapped on zero-volume 2-cycles on a K3
to F-theory and del Pezzo surfaces $B_k$ and that this could reveal the mystery of \cite{inv}.
This should be done by generalizing it in a wider sense than 
that concerning the correspondence between $E_8$ small heterotic instantons and 
LST on $\frac{1}{2}K3=B_9$. Notice that we can analyze this case also 
by considering the 5-brane to be wrapped on $B_9\times\Sigma$.
The intersection matrix on $B_9$ can be given in the form (see \cite{Estrings})
$Q=\left(\matrix{1 & 0\cr 0 & -1}\right)\otimes C_{E_8}$, 
where $C_{E_8}$ is the $E_8$ Cartan matrix, and we can readily calculate the multiloop 
contribution 
$$
\Theta(Z^0)=\Theta_{\Gamma_{1,1}}(\Omega)\times\Theta_{E_8}(\Omega)
$$
which generalizes the amplitude found in \cite{Estrings}
after contraction of the canonical class (dual to the size of the F-theory elliptic fiber)
and the inclusion of the pointlike instanton contribution as given in Section 2.

Moreover in this temptative picture the U-duality group of M-theory on $T^k$ would get
mapped to global diffeomorphisms on the 5-brane wrapped on the del Pezzo.

\section{Conclusions and Open Questions}

Let us start this final section with some more speculative remarks about Little String Theory.

Since its existence has been conjectured it was looked at some completely intractable theory
and still it seems to be. Its properties are very peculiar. It has to be a self-dual superstring
in six dimensions (the only dimension where strings can be self dual) and its low energy 
spectrum has to consist of the self-dual tensor multiplet with respect to the $(0,2)$
target space supersymmetry.
We notice that, because of the six dimensional Dirac quantization condition
for the charge of a self dual string $e^2=\pi n$,
the elementary charge is not an adjustable parameter and therefore,
assuming that the charge is a non constant function of the string coupling,
then it would seem natural to expect that Little String Theory does not admit 
a perturbative world-sheet formulation.
Moreover, as far as we understand, self-duality means that if the string is 
minimally coupled to a 2-form potential as
$$
S\sim \int_\Sigma B +\dots
$$
then, under a world-sheet variation $\Sigma\to \Sigma+\delta\Sigma$, we have
$$
\delta S\sim \int_\Omega dB +\dots
$$
where $\partial\Omega=(\Sigma+\delta\Sigma)_+\cup (\Sigma)_-$ and the lower signs describe the 
relative orientations.
Now, choosing a metric -- or better a conformal class of metrics -- on the target space
we can split $dB=(dB)_++(dB)_-$ in self dual and anti-self dual parts
and claim that really the above variation is
$$
\delta S\sim \int_\Omega (dB)_+ +\dots
$$
and the anti-self dual part is decoupled being a null operator. 
This decoupling statement is the self-duality 
of the string theory from the world-sheet point of view and, 
because of the fact that the possible supersymmetry multiplets in 
six dimensions only allow an anti-self dual field to be part of a
gravitational multiplet,
then it follows that the string has also to be decoupled with respect to the 
six dimensional 
metric and in particular the string should be tensionless
\footnote{
It appeared recently an interesting paper \cite{andreas2} where the world-sheet
coupled theory is studied. Although we don't have a clear link of these
results with ours, let us note that requiring six dimensional conformal
invariance means that, since the scalar fields $\phi^a$ in the tensor
multiplet scale inversly than the metric,
there exists a natural world-sheet coupling term
$\int_\Sigma \sqrt{\phi^a\phi^a} G_{\mu\nu}\partial X^\mu\bar\partial X^\nu$
which might do the job of inducing tension to the strings while the 5-branes 
get far from each other.}.
All this might sound at this point a contradiction: the self duality condition requires the 
introduction of a metric to be defined, but then it has to be decoupled from the string
because of supersymmetry.
These arguments seem to clash with the idea of having a world-sheet description for LST.

Let us notice anyhow that in principle nothing prevents the existence of peculiar string theories
where the string coupling renormalization flow can be followed from a perturbative to a finite 
coupling regime, like the self-dual string seems to require.
Moreover, the peculiar properties of a gravitationally decoupled string theory 
at finite coupling -- if any -- 
can be very different from the properties of string theories we are used to.
In particular notice that the self-duality condition requires the introduction of a conformal class 
of metrics on the six-dimensional manifold and that it is possible that a
conformal symmetry makes it in some sense retractable. This sounds nice from the discussion 
at the end of the last section.

In this paper we have found some potentially interesting links between 5-branes and 
string theories. It seems that we just touched the surface of some deeper connections
and wider duality principles. The multi 5-brane case in particular deserves a more careful 
study and the role played by non-zero characteristics $\theta$-functions and 
SYM theories shifted by $\ZZ_2$ valued cohomologies
should be understood in a geometrical clear way.
Moreover in section 4 we have found the multiloop heterotic lattice summation
with vanishing Wilson lines and $B$ field and it would be interesting to
generalize these results to enclose non zero values for those parameters.
This problem seems to be the counterpart of the vanishing of the $C$ field 
found in \cite{inv}.
Another point to be understood regards a careful study of the modular
properties of possible complete amplitudes in order to see if there are some
potential anomalies and how to cure them by an eventual readjustment 
of the temptative picture that we propose.
Furthermore, a generalization of the results obtained for the heterotic string 
seems available also for the closed type II superstring amplitudes by considering 
the 5-brane wrapped on $\Sigma\times V_k$, where $V_k$ is obtained by a connected 
sum of $k$ copies of $\CP^1\times\CP^1$ and the intersection form being 
$\left(\matrix{0 & 1\cr 1 & 0\cr}\right)^{\oplus k}$.

As we have noticed in the last section, LST seems to give a possibility to understand 
the geometric duality noticed in \cite{inv}.
The more general fact that it contains multiloop $\theta$-functions (and therefore
superstring theory's Hilbert space in its spectrum), makes 
all the story about this duality even more interesting from the LST point of view.
What seems one of the most promising phenomenon in this picture is the changing
role between what is world-sheet and what is target space in non perturbative string 
theory. 
In particular the results in this papaer suggest a possible unified picture including
string world-sheet and target space data in the 5-brane world-volume by which a common origin 
of different discrete symmetries in superstring theory -- namely modularity and (a subgroup of)
U-duality -- is indicated.
This could open new possibilities to understand a 
dynamical generation mechanism for space-time in string theory.

\vspace{.5 cm}
{\bf Acknowledgments}
I would like to thank X.~Bekaert, M.~Bertolini, N.~Boulanger, M.~Matone,
J.F.~Morales, A.~Van Proeyen, A.~Tomasiello and M.~Tonin
for interesting discussions.
Work supported by the European Commission RTN programme
HPRN-CT-2000-00131 
to which the author is associated from Leuven University.

\appendix

\section{The period matrix for factorized geometries}

In this appendix we derive a formula for the period matrix of 6-manifold of product
form $\Sigma_2\times M_4$, where $\Sigma_2$ and $M_4$ are connected manifolds
of dimension 2 and 4 respectively.
The general definition of the period matrix goes as follows.
Let $\left\{E^{(6)},\tilde E^{(6)}\right\}$ be a symplectic basis
of harmonic 3-forms on the six-manifold at hand such that, in matrix notation,
$$
\int E^{(6)}E^{(6)}=0,\quad
\int \tilde E^{(6)}E^{(6)}=1,\quad
\int \tilde E^{(6)}\tilde E^{(6)}=0.
$$
We can expand $\tilde E^{(6)}=X^0E^{(6)}+Y^0{}^*E^{(6)}$, where ${}^*$ is the 
Hodge operator. Then $Z^0$ is defined as $Z^0=X^0+iY^0$.
  
In our case the world volume is in the product form $\Sigma_2\times M_4$
and therefore we have
\be
H^3(\Sigma_2\times M_4)=
H^0(\Sigma_2)\otimes H^3(M_4)\oplus 
H^1(\Sigma_2)\otimes H^2(M_4)\oplus
H^2(\Sigma_2)\otimes H^1(M_4)
\label{fact}\ee
The action of the Hodge operator exchanges the first and the third addenda (which are
in fact isomorphic) while the second one is left invariant.

This means that the period matrix is split in two 
block-diagonal parts. One can be calculated assuming 
$H^1(\Sigma_2)=0$ and the second by assuming $H^1(M_4)=0$
and then the total period matrix is obtained by a direct sum of the two blocks
$$
Z^0=Z^0_I\oplus Z^0_{II}.
$$

To calculate the first factor we write
$\left\{E^{(6)},\tilde E^{(6)}\right\}$ as
$$
E^{(6)}={}^* \chi
\quad{\rm and}\quad
\tilde E^{(6)}=\zeta\wedge\chi
$$
where $\zeta$ is a volume form on $\Sigma_2$ normalized as $\int_{\Sigma_2}\zeta=1$
and $\chi$ is a basis of $H^1(M_4)$ normalized to be orthonormal as 
$\int_{M_4} \chi\wedge {}^*\chi=\1_{b_1}$.
The calculation is straightforward and we obtain the first factor to be
$$
Z^0_I=i\1_{b_1}
$$
where $b_1={\rm dim}H^1(M_4,\IR)$.

For the calculation of the second factor,
we can expand $\left\{E^{(6)},\tilde E^{(6)}\right\}$ in terms of a
symplectic basis $\left\{[a],[b]\right\}$ for $H_1(\Sigma)$, where
$$
\int_{\Sigma} [a][a]=0,\quad
\int_{\Sigma} [a][b]=1,\quad
\int_{\Sigma} [b][b]=0,
$$
and an orthonormal basis $\{e^{(4)}\}$ for $H_2(M_4)$, i.e. 
$\int_{M_4} {}^*e^{(4)}e^{(4)}=1$.
In terms of the previous objects we have
$$
E^{(6)}=e^{(4)}\otimes[b]\quad {\rm and}\quad
\tilde E^{(6)}=Qe^{(4)}\otimes[a]
$$     
where $Q$ is the intersection matrix
on $M_4$ given by $Q=\int_{M_4} e^{(4)}e^{(4)}$.
We calculate ${}^*E^{(6)}=-Qe^{(4)}\otimes{}^*[b]$,
where ${}^*[b]$ is in the two dimensional sense.
By recalling the relation
$$[a]=-\Omega^{(1)}[b]-
\Omega^{(2)}{}^*[b]$$
which holds on every Riemann surface
(see \cite{fk} p. 61-63)
with period matrix 
$\Omega=\Omega^{(1)}+i\Omega^{(2)}$ and the property $Q^2=1$, we get
$$
\tilde E^{(6)}=
\left(-Q\otimes \Omega^{(1)}\right) E^{(6)} +
\left(\1_{b_2}\otimes \Omega^{(2)}\right) {}^*E^{(6)}\, .
$$
Comparing with the general definition we finally read
$$
Z^0_{II}=-Q\otimes\Omega^{(1)}+i\1_{b_2}\otimes \Omega^{(2)}\, ,
$$
where $b_2={\rm dim}H^2(M_4,\IR)$.

The direct sum of the two blocks gives the total period matrix
for product topologies
\be
Z^0=Z^0_I\oplus Z^0_{II}=
i\1_{b_1}\oplus \left\{
-Q\otimes\Omega^{(1)}+i\1_{b_2}\otimes \Omega^{(2)}
\right\}
\la{perma}\ee

\end{document}